\documentclass[preprint]{aastex}
\shorttitle{S.S. HASAN, W. KALKOFEN \& A. A. VAN BALLEGOOIJEN }
\shortauthors{EXCITATION OF OSCILLATIONS IN THE MAGNETIC  NETWORK }
\begin{document}
\title{EXCITATION OF OSCILLATIONS IN THE MAGNETIC  NETWORK ON THE SUN}
\author{\sc S. S. Hasan\altaffilmark{1,2}, W. Kalkofen\altaffilmark{2} 
and A. A.  van Ballegooijen\altaffilmark{2} }
%\affil{\Large\tt shasan@cfa.harvard.edu}
%\affil{\Large\tt wolf@cfa.harvard.edu}
%\affil{\Large\tt avanballegooijen@cfa.harvard.edu}
%\email{hasan@iiap.ernet.in}
\altaffiltext{1}{Indian Institute of Astrophysics, Bangalore 560034, India 
(hasan@iiap.ernet.in)}
\altaffiltext{2}
{Harvard-Smithsonian Center for Astrophysics, 60 Garden
St., Cambridge, MA~02138, U.S.A.}

\begin{abstract}
We examine  the excitation of  oscillations in the magnetic network of
the  Sun through the footpoint   motion of photospheric magnetic flux
tubes located in intergranular lanes.   The motion is derived from
a  time series  of high-resolution  G band  and continuum  filtergrams
using an object-tracking technique.  We model the response of the flux
tube  to the footpoint motion in  terms of the Klein-Gordon equation,
which   is  solved analytically as     an  initial value problem   for
transverse  (kink) waves.  We compute  the  wave energy flux in upward
propagating transverse waves. In general we find that the injection of
energy into  the chromosphere  occurs in  short-duration pulses,
which would lead to a  time variability in chromospheric emission that
is incompatible with observations.  Therefore, we consider the effects of
turbulent convective flows on flux tubes in intergranular lanes. The turbulent
flows are simulated by adding high-frequency motions  (periods   5-50 s) with
an amplitude of 1~km~ s$^{-1}$. The  latter are simulated by adding random
velocity fluctuations to the  observationally determined  velocities.  In  this
case we find that the energy flux is much less intermittent and can in
principle carry adequate energy for chromospheric heating.
\end{abstract}
\newpage
\section{INTRODUCTION}
It is now well established  that the solar photosphere  is threaded  with strong
(kilogauss) magnetic fields in the form of vertical  flux tubes located in the
network, where they are  observed as bright points.  Observations have
revealed  that  these  network bright points  (NBPs)  are  in a highly
dynamical state due  to the buffeting action  of granules (e.g. Muller
1983; Muller et  al. 1994; Berger \& Title 1996; Berger et al. 1998; 
van Ballegooijen  et al. 1998).

The jostling of magnetic elements by granules can excite oscillations in  flux
tubes, which  can in principle  be an important mechanism for chromospheric and
coronal heating, provided that the motions are rapid enough  (Choudhuri,
Auffret  \& Priest 1993;    Steiner, Kn\"olker  \& Sch\"ussler 1994; Hasan \&
Kalkofen 1999,  hereafter  Paper  I). As transverse  waves, excited  in the
photosphere,  travel upwards their velocity amplitude  increases. In
chromospheric layers, the  velocity amplitude becomes comparable to the tube
speed  for kink waves, leading to  an  efficient coupling with  longitudinal
waves (Kalkofen 1997).  The latter  can dissipate  by forming   shocks 
(e.g. Zhugzhda, Bromm \& Ulmschneider 1995).

In Paper I it was argued that  network oscillations can be efficiently
excited through the buffeting action on flux tubes by external granules
with intermittent motions (of about 2 km s$^{-1}$) occurring on a  time
scale of less than about half the cutoff period of kink waves. Typically,
a period of 1 min. is consistent with the general result.
We assumed an analytic form for the external
motions. In the   present work, we use observational data obtained  at
the Swedish Solar Observatory   at La Palma during 1995 (Berger et al.
1998) to infer the horizontal footpoint motions of
magnetic elements in the photosphere.  We model the response of flux
tubes to these motions and calculate the energy  flux in 
upward-propagating transverse (kink) waves. We also simulate the effect  of
hypothetical motions on time  scales of  a few seconds, which  could
significantly  enhance the energy  flux in kink waves.  Finally, we
examine some implications of our results.

\section {MEASUREMENT OF BRIGHT POINT MOTION}
The data were collected on October 5, 1995  between 10:57 and 12:08 UT,
simultaneously  in the  G band  (4305  \AA) and the nearby continuum (4686
\AA), using the phase-diversity method (see L\"ofdahl et al. 1998; 
Berger et al. 1998).  After filtering out  $p-$mode oscillations,  
a time  series of 180 images was obtained,  with  high  spatial  resolution and 
with an interval between frames of 23.5~s. Following Berger et al.,
we  subtract the  images in  the G band and   4686~\AA, which  clearly
highlights the network bright points.

The  motion of   the  bright  points was  followed    using a tracking
technique with ``corks'' as tracers of bright points (van Ballegooijen
et al.  1998, hereafter  Paper II).  The  corks, each with a radius of
60  km (the pixel size of  the current data set),  were advected by an
artificial flow field that is   proportional  to the gradient of   the
intensity in  the magnetic image, and  their positions were followed in time
(for details see Paper II). In Figure 1 we  show the cork positions superposed on
a typical frame of a  magnetic difference image, where $x$ and $y$ denote the
horizontal coordinates on the Sun.  Figures 2a and 2b show the $x$ and $y$
components of the velocity as a function of time for  two representative corks,
whose initial locations are indicated by arrows in Figure 1.

\section {KINK WAVE EXCITATION DUE TO FOOTPOINT MOTION}

\def\k{\kappa}       \def\p{\perp}   \def\b{\beta}      \def\a{\alpha}
\def\g{\gamma}  \def\V{{\cal  V}} \def\Kv{{K$_{\rm  2v}$}} \def\e{{\rm
e}}  \def\d{{\rm  d}}   \def\ratio#1#2{{{#1}\over{#2}}}  \def\e#1{{\rm
e}^{#1}}      \def\ex#1{{e}^{#1}}    \def\ei#1{{\rm  e}^{{\rm   i}#1}}
\def\inv#1{\ratio{1}{#1}}          \def\first#1#2{{{\d#1}\over{\d#2}}}
\def\firstp#1{{\partial\over{\partial#1}}}
\def\firstpp#1#2{{{\partial#1}\over{\partial#2}}}

Let us assume that  the motion  of the G  band  bright points can  be taken as
a proxy for the footpoint  motion of flux tubes at the base of
the photosphere ($z=0$). Furthermore, for  simplicity, we assume that 
flux tubes are isothermal   and ``thin'' compared to the pressure scale height
$H$.  It  is convenient to   work  in terms  of  the  ``reduced'' transverse
displacement $Q_\perp(z,t)=\xi_\perp  \ex{-z/4H}$,   where $\xi_\perp$  is  the
Lagrangian  displacement.  Once the  footpoint motion is  specified, the
velocity ($\dot Q_\perp$) at any height $z$ and time $t$ can  be determined by the following
expression (Paper I):
\begin{equation}
\dot     Q_\perp(z,t)=  \dot    Q_\perp(0,t-z/c_\k)   -k_\k
z\int_0^{t-z/c_\k}              \dot        Q_\perp(0,t_0)
\ratio{J_1(\omega_\k\zeta_\k)}{\zeta_\k}\ \d t_0\ ,
\end{equation}
where $J_1$ is a Bessel function, $\zeta_\k=\sqrt{(t-t_0)^2
-(z/c_\k)^2},$ $c_\k$ and $\omega_\k$ are, respectively, the propagation
speed and cutoff frequency for kink waves, given by:
%\begin{figure}[t]
%\epsscale{0.5} \plotone{flux.ps}
%\caption{Variation with time of the vertical energy flux in transverse
%waves at $z=750$~km due  to footpoint motions corresponding at initial
%locations   (a) (18.7,11.9)~Mm and   (b) (14.5,11.0)~Mm.  The left and
%right panels correspond  to motions associated with  the $x $ and $y $
%components of the velocity respectively.}
%\end{figure}

$$ c_\k^2=\ratio{2}{\gamma}\,\, \ratio{c_s^2}{1+2\b}\,,$$

$$\omega_\k^2  = {g\over 8 H}{1\over  1+2\beta};$$  $c_s$ is the sound
speed, $k_\k=\omega_\k/c_\k$,  $\gamma=5/3$, $\beta=8\pi p/B^2$,  $p(z)$ is
the gas  pressure inside the  tube and  $B(z)$ is  the  magnitude of  the
vertical component of the magnetic field on the tube axis.  The
energy flux ($F_\k$) in a single flux tube is given by (Paper I):
\begin{equation} {F}_{\k} = -
\ratio{2p_{\e,0}}{\b+1}\dot Q_\perp  \Biggl(   \firstpp{Q_\perp}{z}
+\inv{4H} Q_\perp \Biggr)\ex{-z/2H},
\end{equation}
where $p_{e,0}$ denotes the external gas pressure at $z=0$. 

Figures 3a and 3b show the vertical energy flux in transverse 
waves versus time at a
height $z=750$~km for two observed magnetic elements located initially
at the coordinates shown above the figure. The left and right panels
correspond to footpoint motions with the $x $ and $y $ components of the
velocity, respectively. 

We find that the injection of energy into the chromosphere takes place
in brief and intermittent bursts, lasting typically 30 s, separated by
longer periods (longer than the time scale for radiative losses in the
chromosphere) with lower energy flux. The peak energy flux into the
chromosphere is as high as  $\sim 10^9$ erg cm$^{-2}$ s$^{-1}$ in a
single flux tube, although the time-averaged flux is $\sim 10^8$ erg
cm$^{-2}$ s$^{-1}$.  This scenario for heating the magnetic network would
produce strongly intermittent chromospheric emission consisting of brief
intense flashes superimposed on generally a very low background.  However, in dense
network regions the observed chromospheric emission is stably present and
exhibits relatively low amplitude, long-period variations (e.g. Dam\'{e},
Gouttebroze and Malherbe 1984; von Uexk\"{u}ll et al. 1989; Lites, Rutten and
Kalkofen 1993).  Therefore, the above scenario cannot readily explain the
observed persistence of emission from dense network regions.  A possible remedy
is to consider the effect of high frequency motions. 

\section {SIMULATING HIGH-FREQUENCY MOTIONS}

The observations of G band bright points  were taken with  a time cadence of
23.5~s.     However, the possibility  of  random  footpoint motions occurring
on a shorter  time scale cannot be  ruled out.  The solar convection has very
high Reynolds number and is expected to be highly turbulent, involving a wide
range of length and time scales. Indeed, spectroscopic observations of the
solar granulation have shown enhanced broadening of spectral lines in
intergranular lanes, indicating enhanced turbulence within these region (e.g.
Nesis et al. 1992, 1994). Magnetic flux tubes interact with these turbulent
flows within the lanes. We speculate that these interactions produce transverse
displacements of magnetic field lines on length scales much less than the flux
tube diameter ($\sim 100$ km), i.e., the magnetic field inside a flux tube has
fine structure on transverse scales $\ell \ll 100$ km. These perturbations
likely generate Alfv\'{e}n waves which propagate upward along the field lines
and dissipate their energy higher up in chromosphere. 

The velocity at small spatial scale can be estimated as follows.  Assuming the
turbulence has a Kolmogoroff spectrum, the velocity of eddies on length scale
$\ell$ is given by $v( \ell ) \sim v_0 ( \ell / \ell_0 )^{1/3}$, where $\ell_0$
is the outer scale of the turbulence, and $v_0$ is the velocity of the largest
eddies. We assume that $\ell_0$ is given by the width of an intergranular lane
as determined from 3D simulations of the solar granulation (e.g., Stein and
Nordlund 1998), and that $v_0$ is the convective velocity predicted by such
simulations: $\ell_0 \sim$ 100 km and $v_0 \sim$ 2.5 km s$^{-1}$. This turbulence
model predicts that $v\sim$~1~km~s$^{-1}$ on a length scale $\ell \sim$ 10 km, i.e.,
there is significant power in motions at this scale.  The time scale of such
motions is $\ell / v \sim$ 10 s, which is significantly shorter than the time
scale for granular buffeting.

In the following we simulate the effect of such high-frequency motions on the
response of the flux tube (similar to the study of longitudinal wave excitation
in flux tubes by Ulmschneider and Musielak 1998). For simplicity we ignore the
small transverse scale of the motions and we describe the "turbulence" in
terms of footpoint displacements of the flux tube.  We do this by
superposing random motions, with a  rms amplitude of   1 km s$^{-1}$  and a
zero mean  value, on the observationally determined velocity. The random
motions vary on a time scale of 2.35 s.   Such high-frequency  motions (if
present) would not have been detected  in the   present observations.  Hence,
it  is not possible  {\it a priori} to   rule them   out.   Results from such
simulations are shown in Figures 4a and 4b for the $x$ component of the
velocity as a function of time for  two representative corks whose initial
locations are indicated above each panel, and in  Figures 5a and 5b for the
corresponding vertical energy wave flux.   We note that the energy flux in
upward-propagating transverse waves is much  larger    and  less intermittent
than in  the case without high-frequency footpoint motions.

The energy flux  in the form of  high-frequency transverse waves appears
adequate to continuously heat the chromosphere.  Taking a filling factor of
10\% at a height $z=750$~km above the photosphere,  the predicted flux is
approximately \hbox{$2\times 10^7$} erg cm$^{-2}$ s$^{-1}$,   which  is
sufficient to balance the   observed radiative  loss of   the  chromospheric
network ($\sim 10^7$ erg cm$^{-2}$ s$^{-1}$, see Model F$^\prime$ in Avrett
1985)

\section{DISCUSSION }
The energy flux in  upward propagating transverse waves has been shown for a
few  representative magnetic elements. We should point out that we carried
out similar calculations for several corks,  before selecting two as being
typical of the large sample that we examined.

We  find that the injection of energy into the  chromosphere  occurs in  brief
and intermittent bursts (lasting typically  30-60 s), separated  by longer
periods (longer than the typical time scale  for radiative loss) with low
energy flux.   The peak energy  flux into the chromosphere  is typically
$10^{9}$ erg cm$^{-2}$ s$^{-1}$ in a single flux tube.  Observationally, such a
scenario  for  heating of the magnetic network on a 1 arc sec scale would lead
to high variability in Ca II K emission, contrary to observations.  However,
the observations of G band bright  points used here do not have the time
resolution and sensitivity necessary to detect motions on shorter time scales.
We simulate such high-frequency motions by adding random velocity fluctuations
with a zero  mean value and  an rms amplitude of 1~km~s$^{-1}$.  We  conclude
that  for   transverse waves  to provide a  viable mechanism  for  {\it
sustained}   chromospheric   heating, the  main contribution  to the  heating
must come   from high-frequency motions (periods 5-50 s). 

The proposed high-frequency motions have very small spatial scale (10 - 100 km)
and the transverse displacements involved are also very small (for example, a
motion with period 50 s and velocity amplitude of 1~km~s$^{-1}$ has a displacement
amplitude of only 8 km). Such length scales are well below the angular
resolution of present-day solar telescopes.  Therefore, it is unlikely that
such high-frequency motions will ever be detected in studies of the proper
motions of photospheric flux tubes. Spectroscopic methods, such as measurements
of spectral line broadening by unresolved turbulent flows, offer greater
promise but will still require very high angular resolution.

\acknowledgements {The observational data used in this study were provided by
R.A. Shine of Lockheed-Martin and were obtained at the Swedish Vacuum Solar
Telescope.  We acknowledge support from the National Science Foundation and
NASA. SSH is thankful to the Smithsonian Institution for travel support under
SMIN grant no. 9411FR00007. We are grateful to the referee for helpful
comments.}

\newpage
\begin{figure}
\plotone{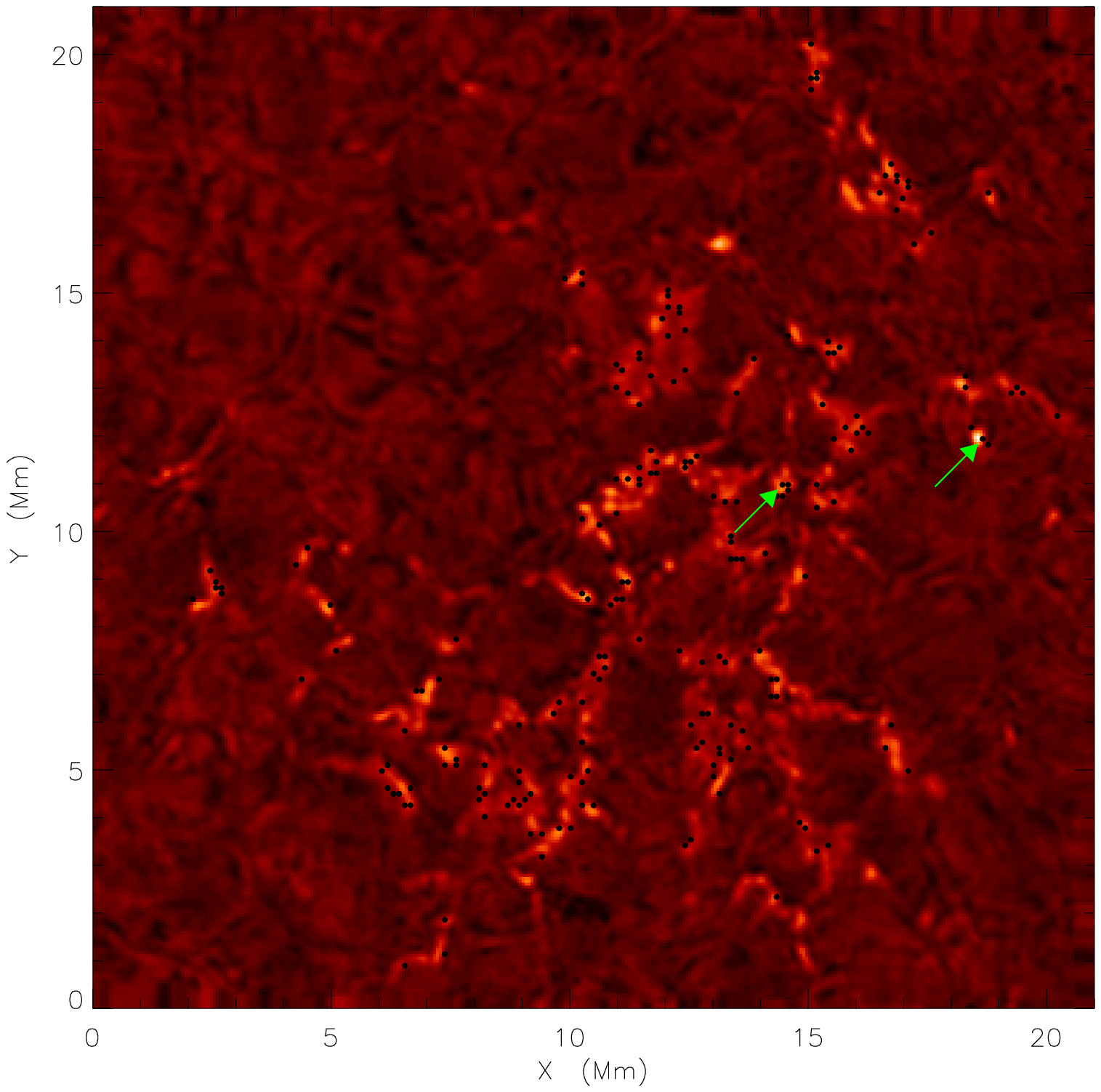}
\caption{   Difference  between G band  and  continuum
images in the initial frame.  The black dots depict the corks, used for
tracking the bright points. The arrows denote the locations of two
representative corks.}\label{fig1}
\end{figure}
\begin{figure}
%\plotfiddle{veloc.ps}{6cm}{0}{80}{80}{-250}{-25}
\plotone{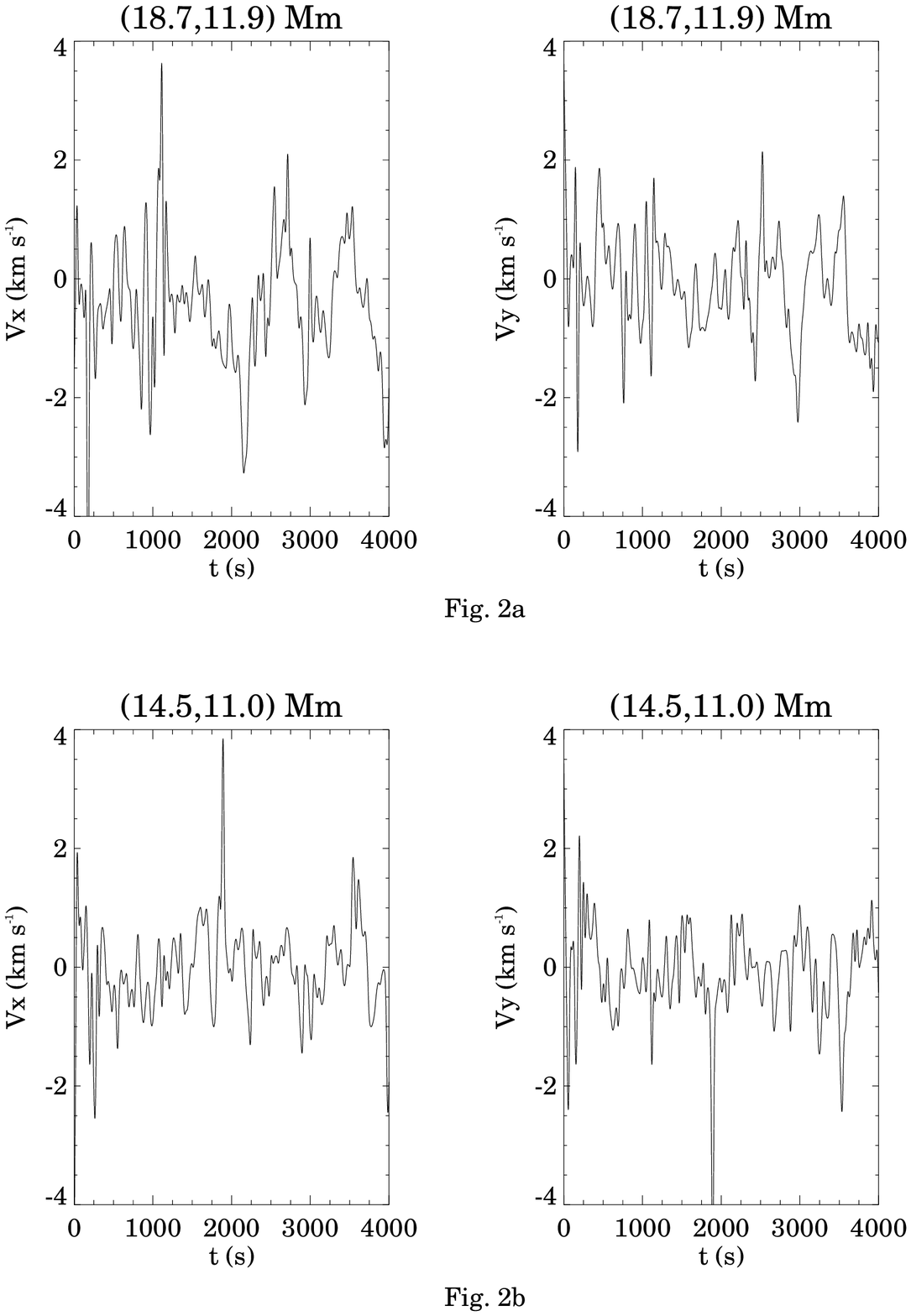}
\caption
{Variation  with time  of the  $x$ and $y$  
components  of the horizontal velocity of G band bright points.   The
coordinates (in Mm) of the corks at the initial instant (shown in Figure  1)
are given above each panel.}  \label{fig2}
\end{figure}
\begin{figure}
%\plotfiddle{flux.ps}{6cm}{0}{80}{80}{-250}{-25}
\plotone{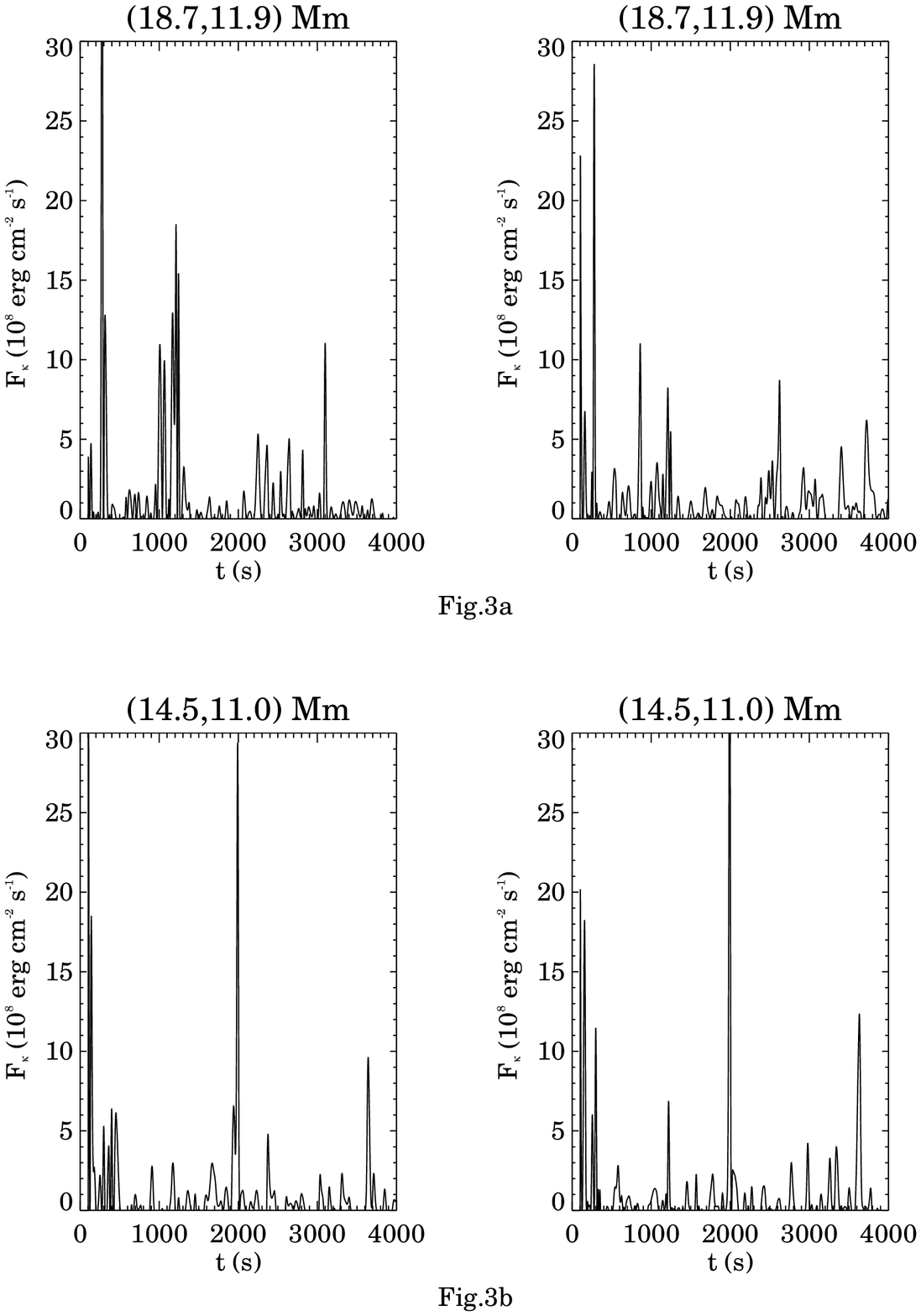}
\caption
{Variation with time of the vertical energy flux in
transverse waves at $z=750$~\hbox{km} due  to footpoint motions
corresponding to initial locations   (a) (18.7~Mm,11.9~Mm) and   (b)
(14.5~Mm,11.0~Mm).  
The left and right panels correspond to motions
associated with the $x $ and $y $ components of the velocity,
respectively.}
\label{fig3}
\end{figure}
\begin{figure}
\plotone{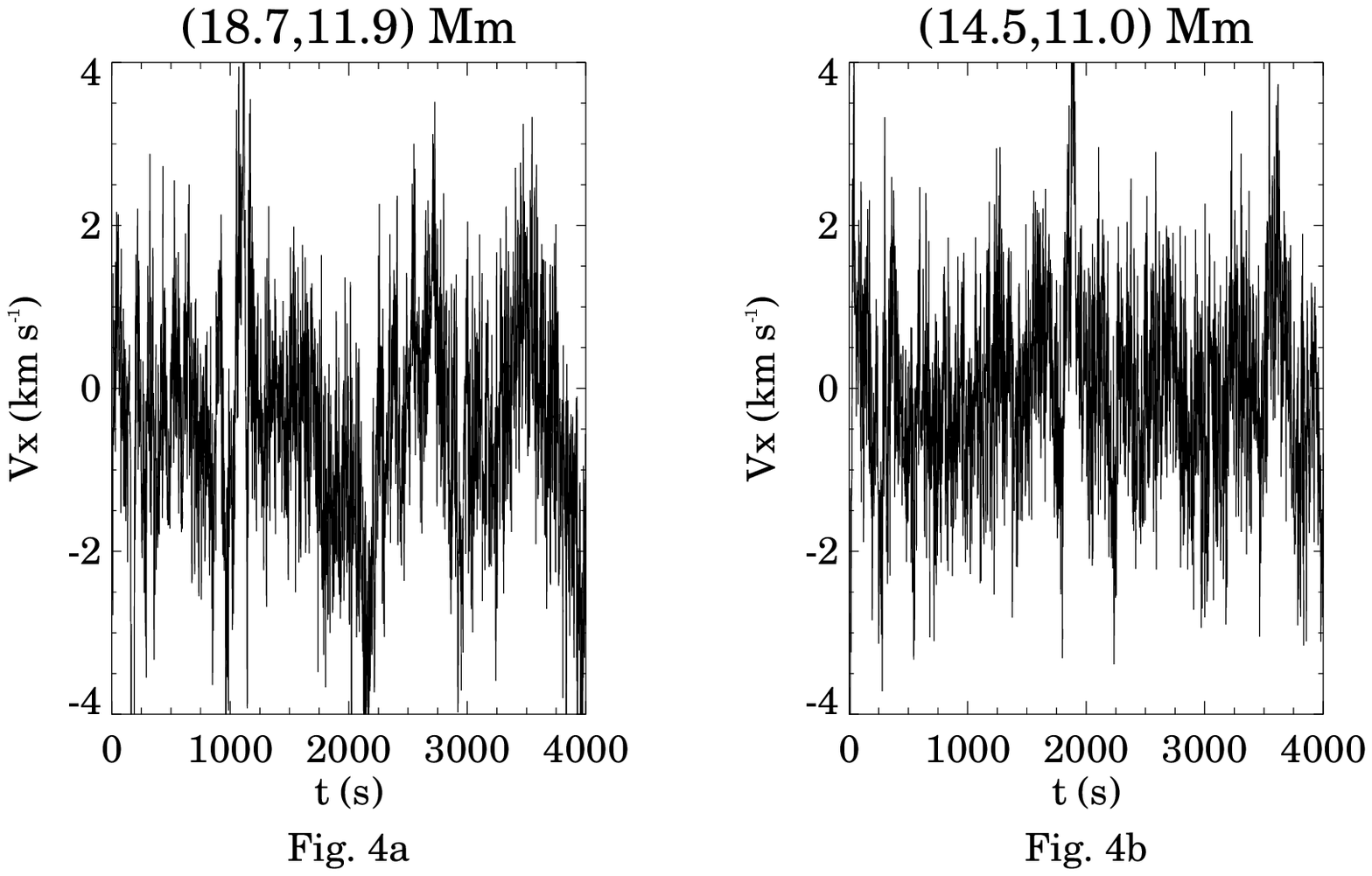}
\caption 
{  Variation with time of   the horizontal velocity
of G band bright points with random noise added for footpoint motions
associated with corks initially at (a) (18.7,11.9) Mm and (b)
(14.5,11.0) Mm.}  
\label{fig4}
\end{figure}
\begin{figure}
\plotone{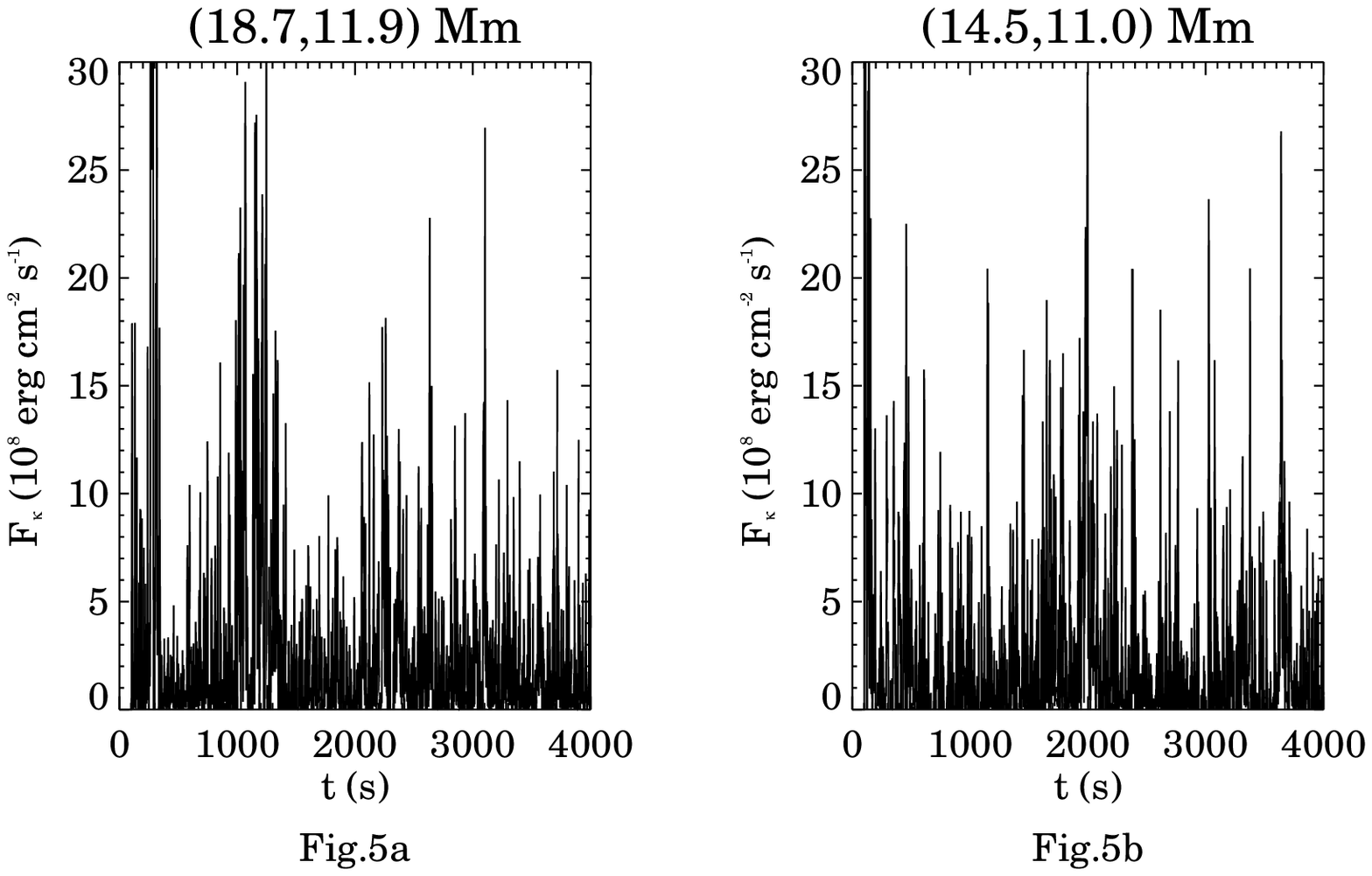}
\caption
{  Variation with  time   of  the   vertical  energy
flux   in transverse waves at  $z=750$~km due to  footpoint motions with
random noise associated  with corks initially at  (a) (18.7,11.9)  Mm
and (b) (14.5,11.0) Mm.}\label{fig5}
\end{figure}
\end{document}